\documentclass[aps,prl,twocolumn,groupedaddress]{revtex4}

\usepackage{graphicx}
\usepackage{amssymb}

\begin{document}


\title{Ordered phase and non-equilibrium fluctuation in stock market}

\author{Jun-ichi Maskawa}
\affiliation{Department of Management Information, Fukuyama Heisei University, Fukuyama, Hiroshima 720-0001, Japan}

% Abstract
\begin{abstract}
We analyze the statistics of daily price change of stock market in the framework of a statistical physics model for the collective fluctuation of stock portfolio. 
In this model the time series of price changes are coded into the sequences of up and down spins, and the Hamiltonian of the system is expressed by spin-spin interactions as in spin glass models of disordered magnetic systems. 
Through the analysis of Dow-Jones industrial portfolio consisting of 30 stock issues by this model, we find a non-equilibrium fluctuation mode on the point slightly below the boundary between ordered and disordered phases. 
The remaining 29 modes are still in disordered phase and well described by Gibbs distribution. 
The variance of the fluctuation is outlined by the theoretical curve and peculiarly large in the non-equilibrium mode compared with those in the other modes remaining in ordinary phase.
\end{abstract}

\maketitle

% main text
%\section{}
%\label{}
The number of days in which prices of all stock issues in Dow-Jones industrial portfolio moved in the same direction is 80 within 3636 trading days in the period from 9/Jul/86 to 22/Nov/00, while its probability is $2^{-29}$ when we assume Bernoulli trials.
How can we explain the factor $10^7$ in the difference between these values?
The methods and the concepts, as scaling and criticality, developed in material science have been applied to the study of financial markets \cite{mandel,mantegna,econo}.
Recently a framework based on spin glass model to study the collective price changes of stock portfolios was proposed \cite{maskawa,takayasu}.
The application to 1-minute price changes in D-J portfolio made clear that the concept of energy works even in financial markets as well as the above physical concepts.
Here we study the properties of daily price changes in D-J portfolio based on the same model and attempt an explanation of the factor $10^7$ by Gibbs factor of canonical distribution.
Through this study we clarify the applicability of the spin glass picture to the price fluctuations in a wide range of time scale and find a novel feature of stock market. That is a non-equilibrium fluctuation mode on the point close to the boundary between ordered and disordered phases. D-J portfolio has been quenched into the unstable region, but does not reach equilibrium. The variance of the fluctuation, which is physically equal to susceptibility and is called (the square of) risk in financial economy, is outlined by the theoretical curve and peculiarly large in the non-equilibrium mode compared with those in the other modes remaining in ordinary phase.

In this paper, we concentrate on the statistics of the sign of price change \cite{maskawa,takayasu}. 
The time series of price changes are coded into the sequences of up and down spins $S_i = \pm 1$ (i=1 to portfolio size N) and the Hamiltonian is expressed by long-range spin-spin interactions as Sherrington-Kirkpatrick model of spin glass \cite{s-k}, which is given by 
\begin{equation}
H[S,h]=-\sum_{<i,j>}J_{ij}S_i S_j-\sum_i h_i S_i,
\label{hamiltonian}
\end{equation}    
We consider a portfolio as a subset of the whole stock market, and the complement of the subset works as heat reservoir. Various observable quantities are obtained as the statistical averages with Gibbs weight assigned to each configuration in equilibrium. The interaction coefficients $J_{ij}$ are constant \cite{jij} but not fixed yet.
The external field $h_i$ is set to be zero in the analysis of actual data \cite{external}.
We use Thouless-Anderson-Palmer (TAP) mean field theory \cite{tap,mezard,fischer} whose fundamental equation is given by
\begin{equation}
m_i=\tanh (\sum_{j}J_{ij}m_j + h_i - \sum_{j}J_{ij}^2(1-m^2_j)m_i).
\label{mi}
\end{equation}    
of $m_i=<S_i>$ to determine those coefficients. The partial differentiation of TAP equation (\ref{mi}) with respect to $h_i$ yields the equation
\begin{equation}
\sum_{k}A_{ik}[{\bf m}]\chi_{kj}=\delta_{ij}
\label{achi}
\end{equation}    
where $\chi_{ij}=\partial m_i/\partial h_j$ is susceptibility and
\begin{equation}
A_{ij}[{\bf m}]= -J_{ij}-2J_{ij}^2m_i m_j +\delta_{ij}[ \sum_{k}J_{ik}^2(1-m^2_k)+\frac{1}{1-m_i^2}].
\label{aij}
\end{equation}    
On the other hand, fluctuation-response theorem relates the susceptibility $\chi_{ij}$ and the covariance $C_{ij}=<(S_i-m_i)(S_j-m_j)>$ as
\begin{equation}
\chi_{ij}=C_{ij}.
\label{chi}
\end{equation}    
Substituting the equation (\ref{chi}) into (\ref{achi}), we can derive the relation between $J_{ij}$ and $C_{ij}$. Interpreting $C_{ij}$ as the time average of empirical data over the observation time, $J_{ij}$ are phenomenologically determined by the equation
\begin{equation}
J_{ij}=\frac{-1+\sqrt{1-8C^{-1}_{ij}m_i m_j}}{4m_i m_j}
\label{jij}
\end{equation}
where  $ C^{-1}_{ij}$ is the (i, j)-element of the inverse of the covariance matrix. In the case with $m_i\approx0$ the approximation $J_{ij}\approx -C^{-1}_{ij}$ is applicable.

\begin{figure}
\includegraphics[width=6cm]{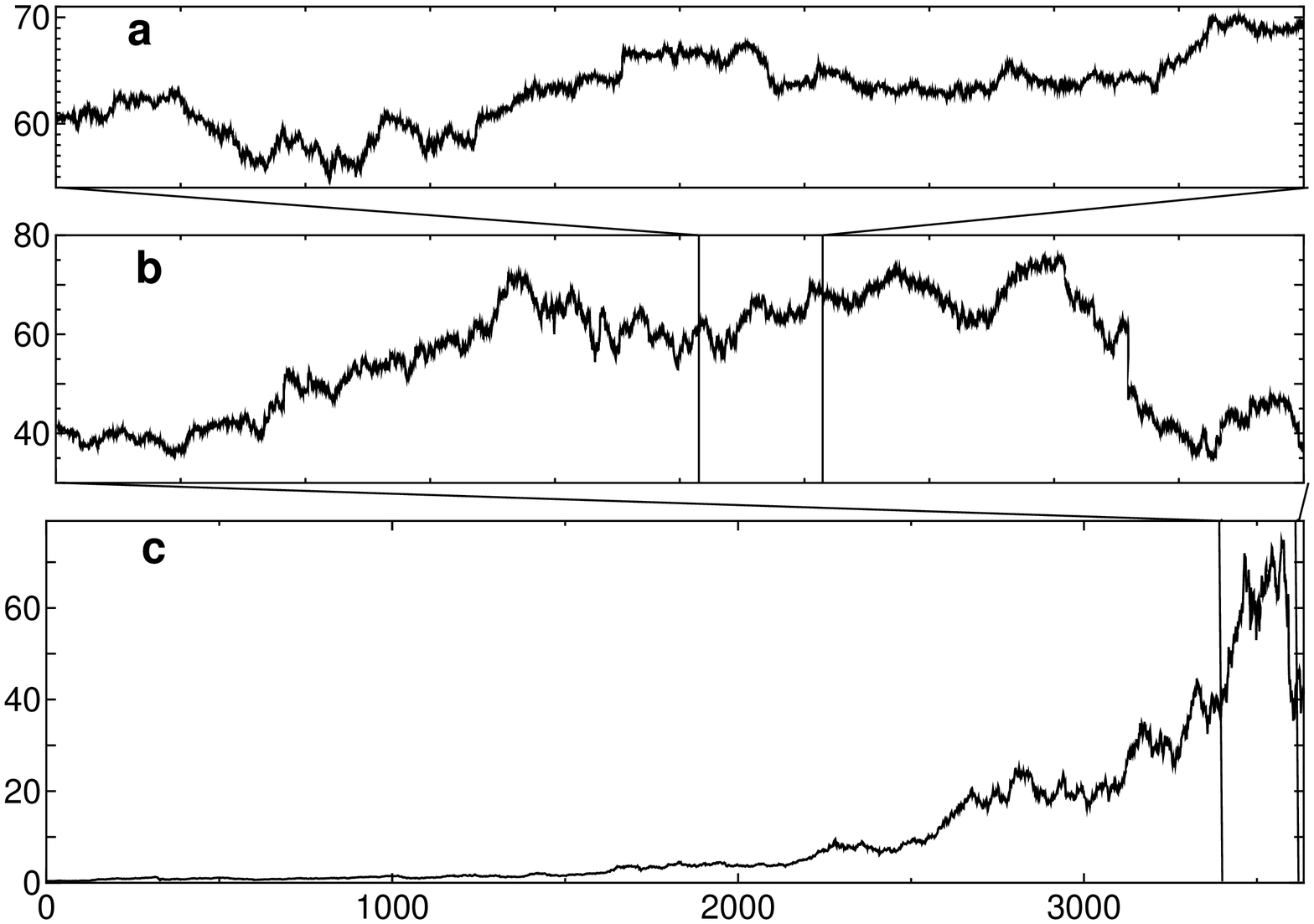}
\caption{Time evolution of Intel's stock price on three time scales. 
{\bf a} The data sampled at 1-minute intervals in the period from 16-May-2000 to 21-Jun-2000.
{\bf b} The data sampled at 10-minute intervals in the period from 1-Dec-1999 to 11-Nov-2000.
{\bf c} The daily data in the period from 9-Jul-1986 to 22-Nov-2000 (3636 days).
}
\label{fig1}
\end{figure}

We investigate three datasets a, b and c of D-J portfolio consisting of N=30 stock issues in the framework of the model. The dataset a is the time series of stock prices sampled at 1-minute intervals in the period from 16-May-2000 to 21-Jun-2000, b is sampled at 10-minutes intervals in the period from 1-Dec-1999 to 11-Nov-2000, and c is the daily price in the period from 9-Jul-86 to 22-Nov-2000. The time evolutions of Intel's stock price on each time scale are shown in Fig. \ref{fig1} for an example. The time series of price changes are coded into the sequences of up and down spins. The covariance $C_{ij}$ for 435 pairs of i and j are derived from the coded data. Then the interaction coefficients $J_{ij}$ are calculated by the equation (\ref{jij}) \cite{ferro}. The energy spectra of the system (portfolio energy) are defined as the eigenvalues of the Hamiltonian $H[S,0]$. The probability density of portfolio energy for each dataset is empirically obtained from the relative frequency during the observation times as
\begin{equation}
p(E) =P(E-\frac{\Delta E}{2} \le H[S,0] \le E+\frac{\Delta E}{2})/\Delta E.
\label{pe}
\end{equation}
In our formalism that is given by the equation $p(E)=n(E)e^{ -E}/Z$ with the density of states $n(E)$ and the partition function $Z$ if the system is in equilibrium. The empirical probability weight $p(E)n(0)/p(0)n(E)$ is plotted in Fig. \ref{fig2} and compared with Gibbs weight as the theoretical prediction. In the panels a and b we see a remarkable fit of equilibrium line to empirical data. From these figures, the price changes sampled at 1 and 10-minutes intervals are well described by the model and are close to equilibrium. On the other hand, in the panel c some deviation of data from theoretical line is observed. Monte Carlo simulation of the probability distribution of the system magnetization $m=(1/N)\sum_{i=1}^{N}S_i$, which describes the degree of the alignment of price changes in portfolio, make more clear this difference in the nature of price changes on the different time scales. The result is shown in Fig. \ref{fig3}. We see the theory explains the empirical data very well in the panels a and b, while the dataset c shows entirely different profile from the theoretical prediction. Does this difference indicate the break down of the theory on the time scale of day? We investigate the theory in more detail by the mode analysis of TAP equation (\ref{mi}), and intend to explain this phenomenon in the framework of the theory.  

The stability of the ordinary phase is analyzed by the linear analysis of the equation (\ref{mi}). In our case, that is
\begin{equation}
\sum_{j} A_{ij}[{\bf 0}]m_j=h_i.
\label{linear}
\end{equation}
The diagonalization of symmetric matrix $A_{ij}$ of the equation (\ref{aij}) solve the equation (\ref{linear}) as $m_{\lambda}=h_{\lambda}/A_{\lambda}$, where $A_{\lambda}$ are eigenvalues of $A_{ij}$, $m_{\lambda}=\sum <\lambda\mid i> m_i$, and $h_{\lambda}=\sum <\lambda\mid i> h_i$ with the real orthogonal eigenvectors $<\lambda\mid i>$. The critical temperature $T_c$ is determined by the equation 
\begin{equation}
A_{\lambda_{min}}=0
\label{amin}
\end{equation}
$A_{\lambda_{min}}$ is the minimum eigenvalue of $A_{ij}$, which correspond to the maximum eigenvalue of the matrix $J_{ij}$ in the linear approximation. However $\beta=1/T$ is included into $J_{ij}$ in our case, so $A_{\lambda_{min}}$ itself fills the role of the measure of the distance from the critical point. In our systems a, b and c, 
$ A_{\lambda_{min}}= $ 0.353, 0.195 and -0.018 respectively. The systems c is in the position slightly below the critical point, while the systems a and b remain in ordinary phase. In order to visualize the statement, TAP free energy \cite{tap} of the system c is shown in Fig. \ref{fig4}, in which the parameter space $\{m_i\}$ is projected on the mode $m_{\lambda}$. In Fig. \ref{fig4}, the upper and lower curves represent TAP free energy as the function of the modes $m_{\lambda}$ belonging to the maximum and the minimum eigenvalues of $A_{ij}$ respectively. The system c is in the unstable region, because the lower curve of the free energy is concave at $ m_{\lambda_{min}}= $-0.067. 
The behavior of the price fluctuation belonging to this mode is entirely different from the remaining 29 modes. The probability of the magnetization fluctuation $S_{\lambda}=\sum <\lambda\mid i> S_i$ of the unstable mode and that of a ordinary mode belonging to the maximum eigenvalue $ A_{\lambda_{max}}$ are plotted in Fig. \ref{fig5} by filled circles and filled boxes respectively. The fluctuation distribution of the modes in ordinary phase are narrow and shows good agreement with the equilibrium distribution, while that of unstable mode is broad and does not reach equilibrium distribution.

The variance of each mode in the system a, b and c is plotted against eigenvalue $A_{\lambda}$ in Fig. \ref{fig6}. Those are compared with the theoretical curve $1/A_{\lambda}$ for infinite equilibrium system. The variance of the fluctuation, which is physically equal to susceptibility and is called (the square of) risk in financial economy, is outlined by the theoretical curve and peculiarly large in the non-equilibrium mode compared with those in the other modes remaining in ordinary phase. Closing to the critical point $A_{\lambda}=0$, however, we find a systematic deviation of data from the theoretical curve, which is due to the non-equilibrium property and the finite size effect of the systems.

The factor $10^7$ problem prompting us to introduce the interaction energy between stocks in the analysis of its collective fluctuation was partially explained by Gibbs weight in our model besides the single non-equilibrium mode. The physical picture of financial markets given here will be useful in the risk management and the selection of portfolios. The applicability of spin glass model to economical system seems to suggest the ubiquity of the applications to other fields such as competitive or frustrated ecological systems.

\bibliography{order}

\begin{thebibliography}{12}
\expandafter\ifx\csname natexlab\endcsname\relax\def\natexlab#1{#1}\fi
\expandafter\ifx\csname bibnamefont\endcsname\relax
  \def\bibnamefont#1{#1}\fi
\expandafter\ifx\csname bibfnamefont\endcsname\relax
  \def\bibfnamefont#1{#1}\fi
\expandafter\ifx\csname citenamefont\endcsname\relax
  \def\citenamefont#1{#1}\fi
\expandafter\ifx\csname url\endcsname\relax
  \def\url#1{\texttt{#1}}\fi
\expandafter\ifx\csname urlprefix\endcsname\relax\def\urlprefix{URL }\fi
\providecommand{\bibinfo}[2]{#2}
\providecommand{\eprint}[2][]{\url{#2}}

\bibitem[{\citenamefont{Mandelbrot}(1982)}]{mandel}
\bibinfo{author}{\bibfnamefont{B.}~\bibnamefont{Mandelbrot}},
  \emph{\bibinfo{title}{The Fractal Geometry of Nature}}
  (\bibinfo{publisher}{W.H.Freeman}, \bibinfo{address}{New York},
  \bibinfo{year}{1982}).

\bibitem[{\citenamefont{Mantegna and Stanley}(1995)}]{mantegna}
\bibinfo{author}{\bibfnamefont{R.}~\bibnamefont{Mantegna}} \bibnamefont{and}
  \bibinfo{author}{\bibfnamefont{H.}~\bibnamefont{Stanley}},
  \bibinfo{journal}{Nature} \textbf{\bibinfo{volume}{376}}, \bibinfo{pages}{46}
  (\bibinfo{year}{1995}).

\bibitem[{\citenamefont{Mantegna and Stanley}(2000)}]{econo}
\bibinfo{author}{\bibfnamefont{R.}~\bibnamefont{Mantegna}} \bibnamefont{and}
  \bibinfo{author}{\bibfnamefont{H.}~\bibnamefont{Stanley}},
  \emph{\bibinfo{title}{An Introduction to EconophysicsCorrelations and
  Complexity in Finance}} (\bibinfo{publisher}{Cambridge University Press},
  \bibinfo{address}{Cambridge}, \bibinfo{year}{2000}).

\bibitem[{\citenamefont{Maskawa}()}]{maskawa}
\bibinfo{author}{\bibfnamefont{J.}~\bibnamefont{Maskawa}},
  \eprint{cond-mat/0011149}.

\bibitem[{\citenamefont{H.Takayasu}(in press)}]{takayasu}
\bibinfo{editor}{\bibnamefont{H.Takayasu}}, ed.,
  \emph{\bibinfo{title}{Empirical Science of Financial Fluctuations}}
  (\bibinfo{publisher}{Springer-Verlag}, \bibinfo{address}{Tokyo},
  \bibinfo{year}{in press}).

\bibitem[{\citenamefont{Sherrington and Kirkpatrick}(1975)}]{s-k}
\bibinfo{author}{\bibfnamefont{D.}~\bibnamefont{Sherrington}} \bibnamefont{and}
  \bibinfo{author}{\bibfnamefont{S.}~\bibnamefont{Kirkpatrick}},
  \bibinfo{journal}{Phys. Rev. Lett.} \textbf{\bibinfo{volume}{35}},
  \bibinfo{pages}{1792} (\bibinfo{year}{1975}).

\bibitem[{jij()}]{jij}
\bibinfo{note}{The interaction coefficients may be dynamical. The decomposition
  of this variable into the dynamical and constant parts is proposed in the
  paper \cite{maskawa}. Even if there are several time scales in the dynamical
  part, they are irrelevant to the statistics under the assumption of Gaussian
  distribution on the fluctuation of the dynamical part. A trial for the
  justification of the assumption was made in the paper.}

\bibitem[{ext()}]{external}
\bibinfo{note}{It is possible to give the economical meaning to the external
  fields $h_{i}$. Examples are policy changes of governments, business
  condition, fundamentals of companies and so on. They act as a trigger for
  price changes from the outside of markets. The condition $h_i=0$ is 0-th
  approximation. The empirical results in this paper show, however, the
  non-zero effect will no more than a small correction at least in our
  framework.}

\bibitem[{\citenamefont{D.J.Thouless et~al.}(1977)\citenamefont{D.J.Thouless,
  P.W.Anderson, and R.G.Palmer}}]{tap}
\bibinfo{author}{\bibnamefont{D.J.Thouless}},
  \bibinfo{author}{\bibnamefont{P.W.Anderson}}, \bibnamefont{and}
  \bibinfo{author}{\bibnamefont{R.G.Palmer}}, \bibinfo{journal}{Phil. Mag.}
  \textbf{\bibinfo{volume}{35}}, \bibinfo{pages}{593} (\bibinfo{year}{1977}).

\bibitem[{\citenamefont{M.Mezard and M.A.Virasoro}(1987)}]{mezard}
\bibinfo{author}{\bibfnamefont{G.}~\bibnamefont{M.Mezard}} \bibnamefont{and}
  \bibinfo{author}{\bibnamefont{M.A.Virasoro}}, \emph{\bibinfo{title}{Spin
  Glass Theory and Beyond}} (\bibinfo{publisher}{World Scientific},
  \bibinfo{address}{Singapore}, \bibinfo{year}{1987}).

\bibitem[{\citenamefont{K.H.Fischer and J.A.Hertz}(1991)}]{fischer}
\bibinfo{author}{\bibnamefont{K.H.Fischer}} \bibnamefont{and}
  \bibinfo{author}{\bibnamefont{J.A.Hertz}}, \emph{\bibinfo{title}{Spin
  Glasses}} (\bibinfo{publisher}{Cambridge University Press},
  \bibinfo{address}{Cambridge}, \bibinfo{year}{1991}).

\bibitem[{fer()}]{ferro}
\bibinfo{note}{In D-J portfolio investigated here, most (92 $\%$ for dataset a,
  b and 85 $\%$ for c) of 435 $J_{ij}$ are positive and the distribution is
  different from Gaussian which is assumed in S-K model. In the dataset c, for
  example, the maximum, the minimum, the mean value and the standard deviation
  of the distribution of $J_{ij}$ are $max=0.299$, $min=-0.044$, $\mu=0.036$
  and $\sigma=0.041$ respectively. The system is rather a random ferromagnet
  than a spin glass.}

\end{thebibliography}

\newpage

\newpage

\begin{figure*}
\includegraphics[width=6cm]{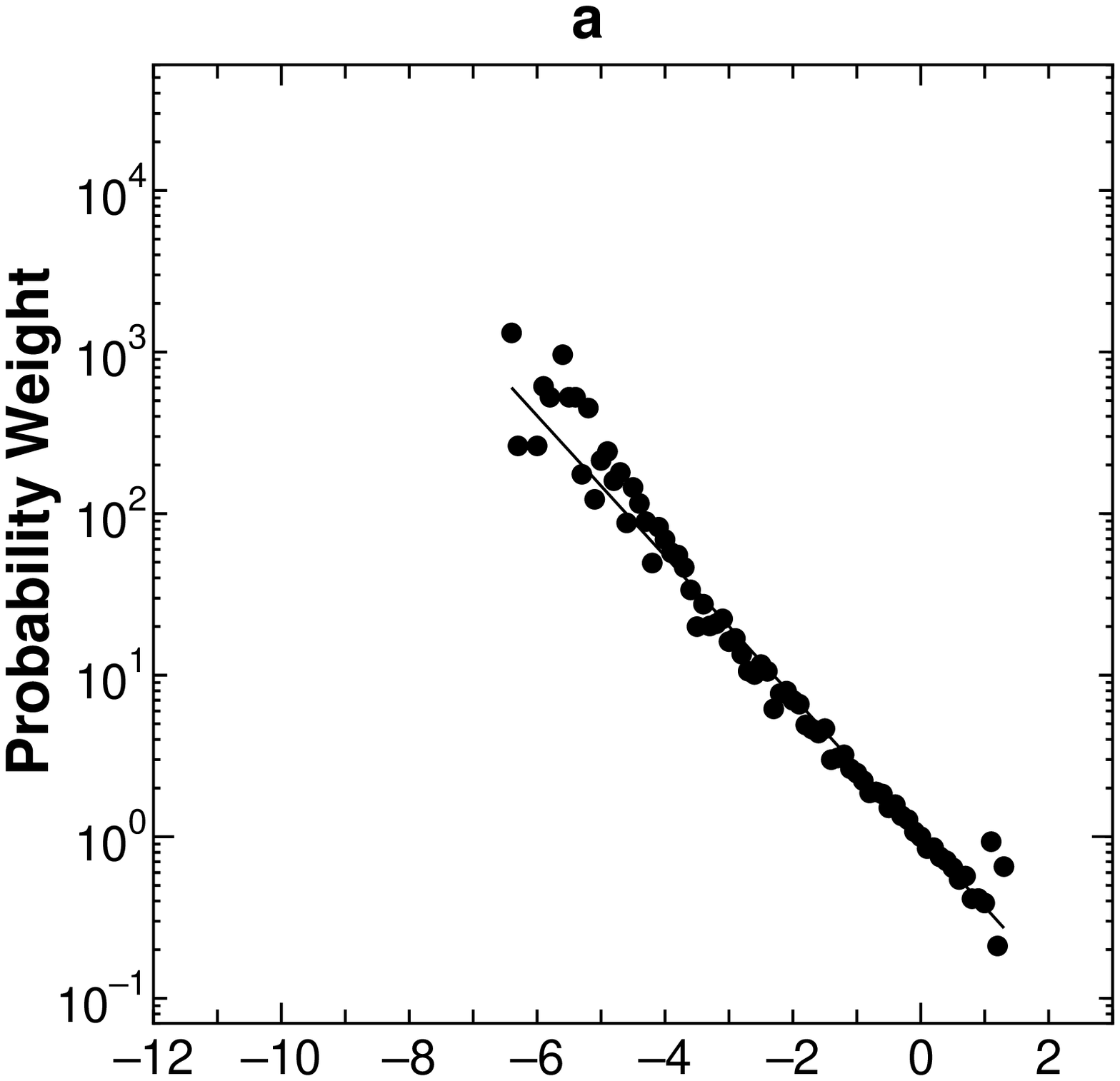}
\end{figure*}
\begin{figure*}
\includegraphics[width=6cm]{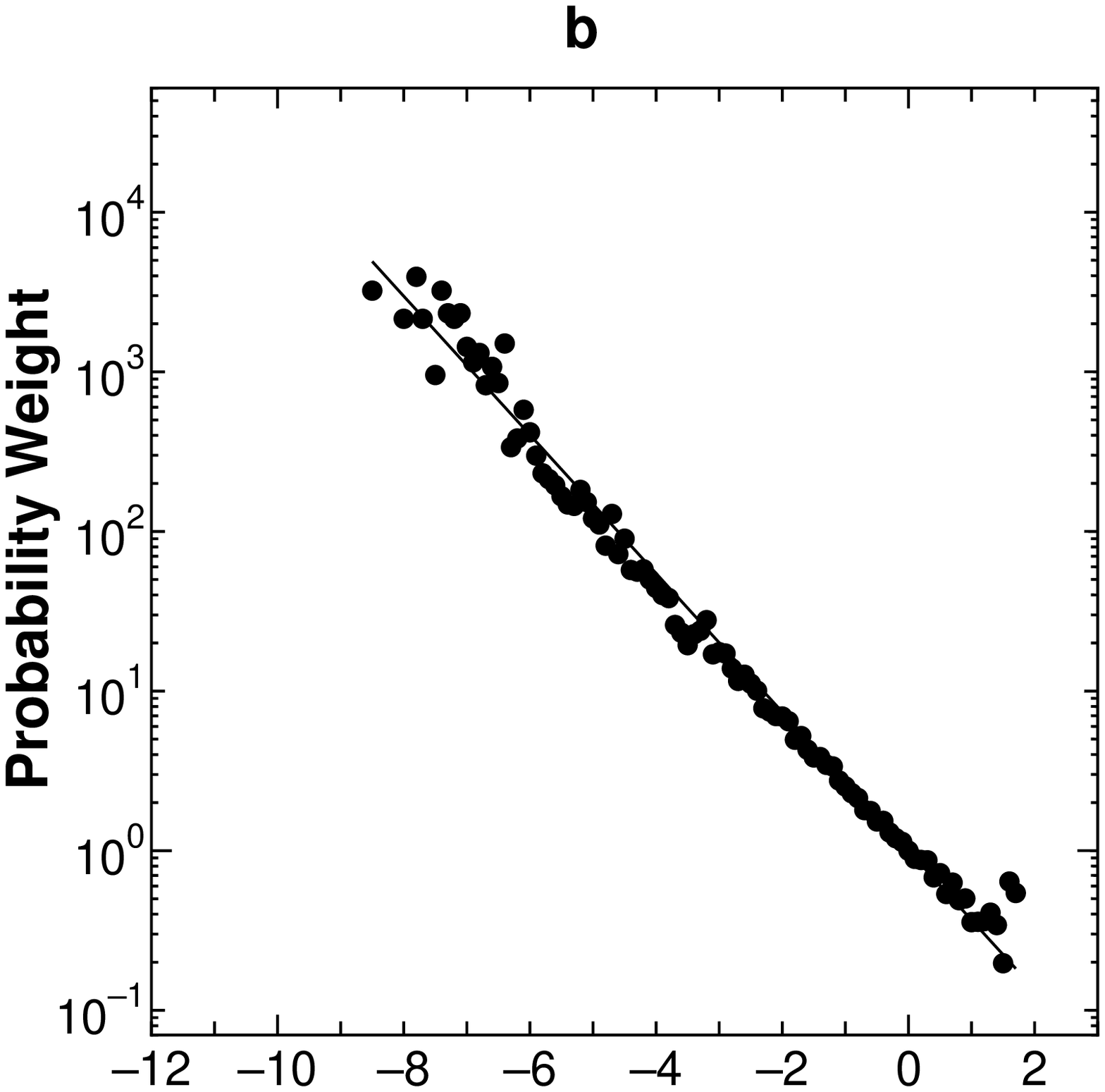}
\end{figure*}

\begin{figure*}
\includegraphics[width=6cm]{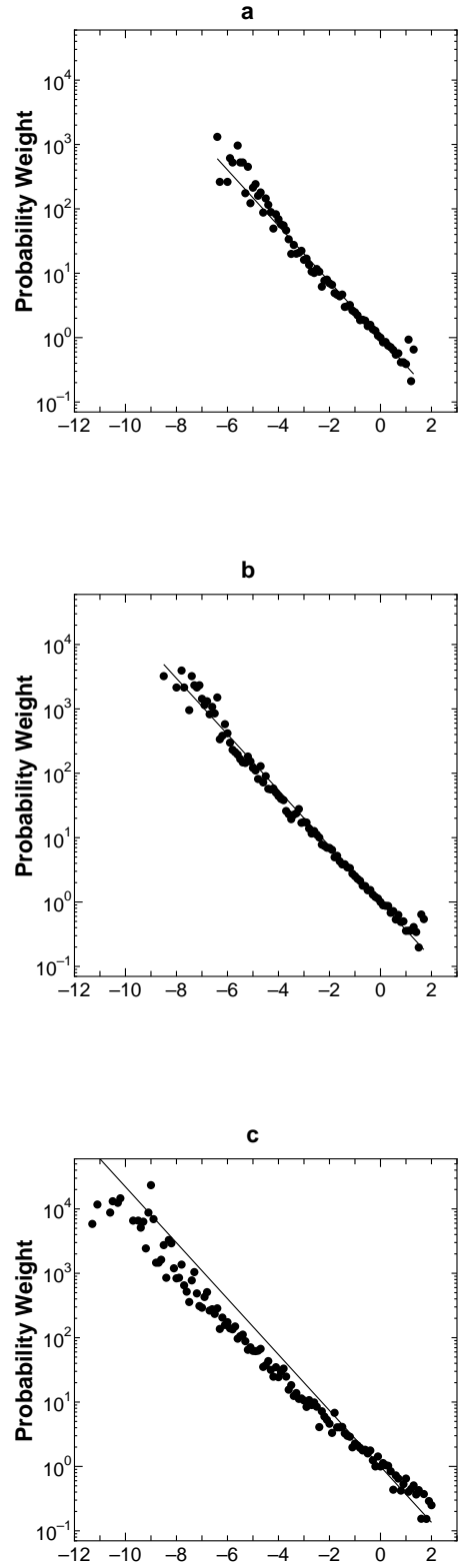}
\caption{Semi-log plot of the probability weight of portfolio energy $E$. 
Filled circle ($\bullet$):the empirical probability weight $p(E)n(0)/p(0)n(E)$.The probability density $p(E)$ is the empirical result, and the density of states $n(E)$ is numerically obtained by $2^{21}$ random sampling from $2^{30}$ configurations. $\Delta E=0.1$.
Solid line: Gibbs weight $e^{-E}$.
{\bf a} The result for the dataset a.
{\bf b} The result for the dataset b.
{\bf c} The result for the dataset c.
}
\label{fig2}
\end{figure*}

\begin{figure*}
\includegraphics[width=6cm]{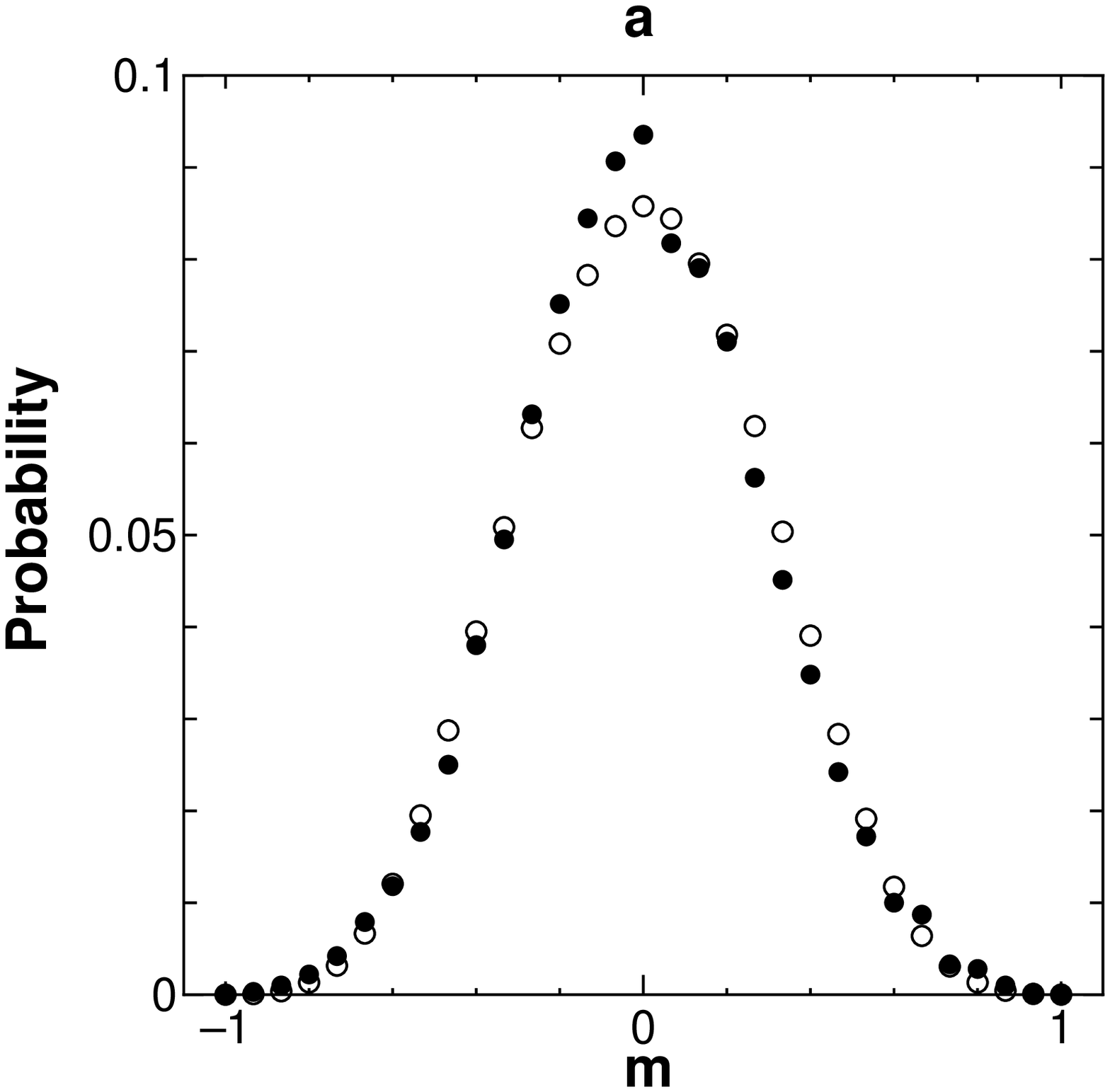}
\end{figure*}
\begin{figure*}
\includegraphics[width=6cm]{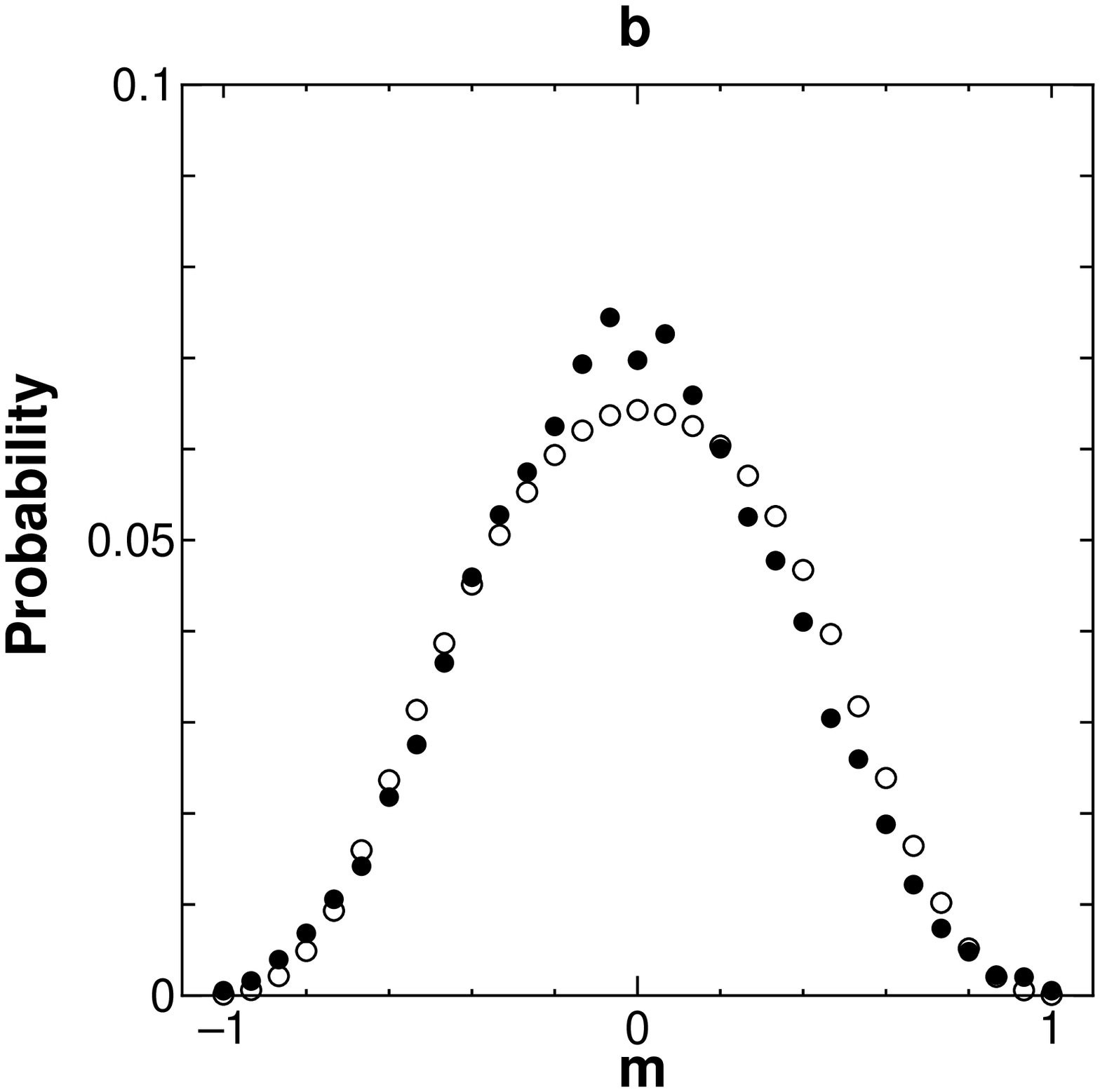}
\end{figure*}

\begin{figure*}
\includegraphics[width=6cm]{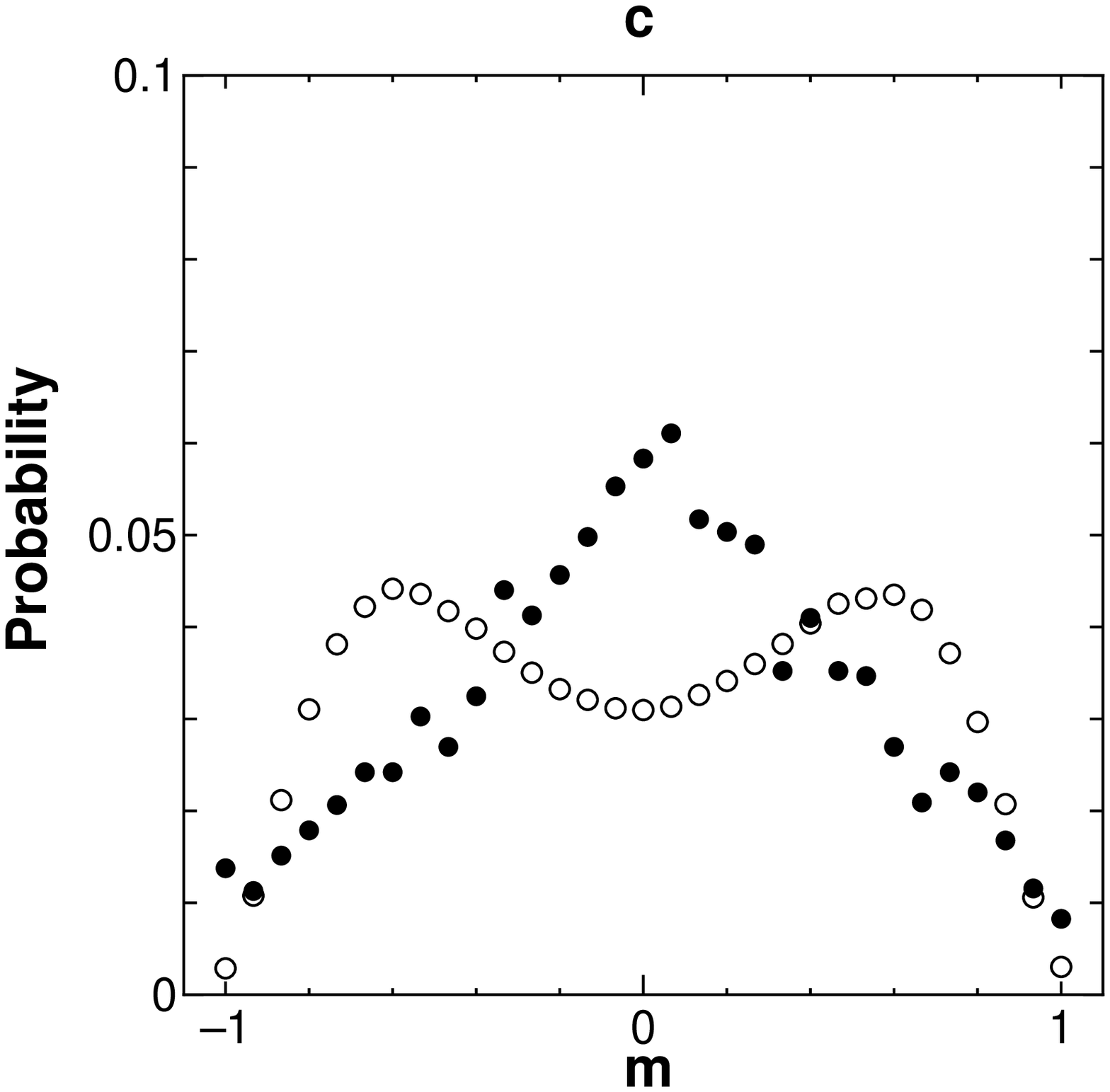}
\caption{Probability of the system magnetization.
Filled circle ($\bullet$):the relative frequency.
Circle ($\circ$): Monte Carlo simulation of $2^{21}$ steps.
{\bf a} The result for the dataset a.
{\bf b} The result for the dataset b.
{\bf c} The result for the dataset c.
}
\label{fig3}
\end{figure*}

\begin{figure*}
\includegraphics[width=6cm]{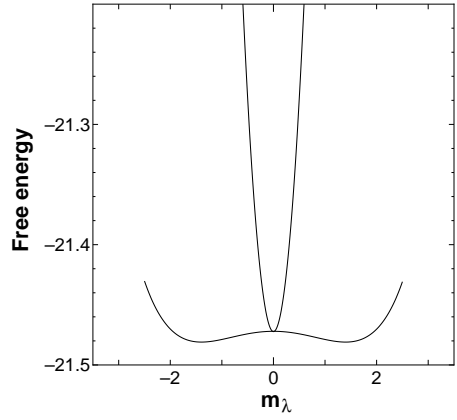}
\caption{Free energy as the function of the magnetization $m_{\lambda}$  for the system c. 
The upper and lower curves represent TAP free energy as the functions of the modes $ m_{\lambda_{max}}=\sum<\lambda_{max}\mid i>m_i$ 
and $m_{\lambda_{min}}=\sum<\lambda_{max}\mid i>m_i$ respectively.
}
\label{fig4}
\end{figure*}

\begin{figure*}
\includegraphics[width=6cm]{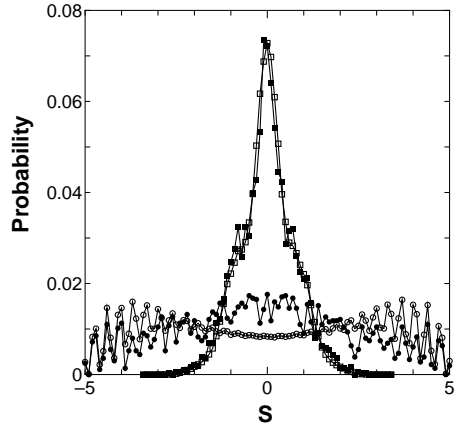}
\caption{Probability of magnetization fluctuation $S_{\lambda}$  for the dataset c.  Filled boxes ($\blacksquare$) and circles ($\bullet$) represent relative frequency
$P(S-\Delta S/2 \le S_{\lambda} \le S+\Delta S/2)$ with $\Delta S=0.1$ for the mode $S_{\lambda_{max}}=\sum<\lambda_{max}\mid i>S_i$ and $S_{\lambda_{min}}=\sum<\lambda_{min}\mid i>S_i$ respectively. Boxes ($\square$) and circles ($\circ$) represent Monte Carlo simulations based on the model.
}
\label{fig5}
\end{figure*}

\begin{figure*}
\includegraphics[width=6cm]{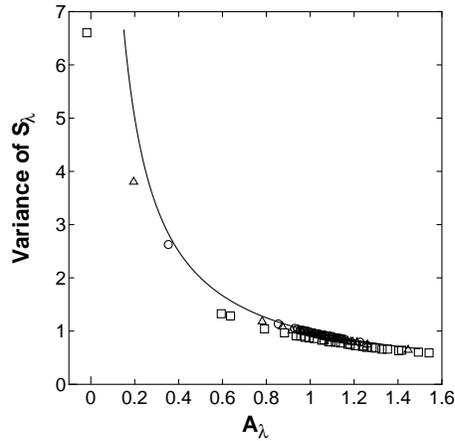}
\caption{Variance of the modes $S_{\lambda}$ against the eigenvalue $A_{\lambda}$. Circles ($\circ$), triangles ($\triangle$), and boxes ($\square$) represent the variance
for the datasets a, b and c respectively. Solid line represents the theoretical curve $1/A_{\lambda}$ for infinite equilibrium system. 
}
\label{fig6}
\end{figure*}

\end{document}